# Effect of Proton Irradiation Temperature on Zinc Oxide Metal-Semiconductor-Metal Ultraviolet Photodetectors

Thomas A. Heuser,[1] Caitlin Chapin,[2] Max Holliday,[1,2] Yongqiang Wong,[3] and Debbie G. Senesky[2,4]

[1]*Department of Materials Science and Engineering, Stanford University, Stanford, California 94305, USA*
[2]*Department of Aeronautics and Astronautics, Stanford University, Stanford, California 94305, USA*
[3]*Department of Electrical Engineering, Stanford University, Stanford, California 94305, USA*
[4]*Material Science and Technology Division, Los Alamos National Laboratory, Los Alamos, New Mexico 87545, USA*

The electrical and structural characteristics of 50 nm zinc oxide (ZnO) metal-semiconductor-metal (MSM) ultraviolet (UV) photodetectors subjected to proton irradiation at different temperatures are reported and compared. We irradiated the devices with 200 keV protons to a fluence of $10^{16}$ cm$^{-2}$. Examination of the X-ray diffraction (XRD) rocking curves indicates a strongly preferred (100) orientation for the grains of the as-deposited film, with decreases in crystal quality for all irradiated samples. In addition, peak shifts in XRD and Raman spectra of the control sample relative to well-known theoretical positions are indicative of tensile strain in the as-deposited ZnO films. We observed shifts of these peaks towards theoretical unstrained positions in the irradiated films relative to the as-deposited film indicate partial relaxation of this strain. Raman spectra also indicate increases of oxygen vacancies ($V_O$) and zinc interstitials ($Zn_i$) relative to the control sample. Additionally, photocurrent versus time measurements showed up to 2x increases in time constants for samples irradiated at lower temperatures months after irradiation, indicating that the defects introduced by suppression of thermally-activated dynamic annealing process has a long-term deleterious effect on device performance.

Electronics intended for use in space environments are subjected to high levels of radiation from a variety of sources over their operational lifetimes, including protons and electrons trapped inside planetary magnetic fields, ultraviolet (UV) radiation, and x-rays emitted from the sun during solar flares and coronal mass ejections, as well as a wide range of light and heavy ions from galactic cosmic rays (GCRs).[1] High-energy radiation can damage exposed devices on a material level, introducing crystalline defects by displacing atoms from lattice sites, thereby degrading device microstructure and electrical characteristics.

Devices made using wide-bandgap semiconductor materials such as zinc oxide (ZnO), gallium nitride (GaN), and silicon carbide (SiC), which have large atomic displacement energies (ZnO: 57 eV, GaN: 19.5 eV, 4H-SiC: 21.3 eV) are much more resistant to radiation-induced degradation than those made with conventional semiconductors like silicon (12.9 eV) or gallium arsenide (9.5 eV).[2,3,4] In particular, ZnO-based devices have demonstrated extreme resilience against a variety of types of radiation, including proton,[5,6] electron,[7] gamma.[8,9] This resistance to radiation damage comes not only from large displacement energies, but also from high rates of dynamic annealing (self-healing of irradiation damage while the irradiation is still occurring), which are enhanced by the high mobility of radiation-induced point defects and the more ionic nature of its interatomic bonds, and grant ZnO self-healing capabilities significantly beyond even other wide-bandgap materials like GaN.[10,11] As dynamic annealing depends on thermally-activated diffusion processes, device temperature during irradiation has a significant impact on self-healing ability.[7,10]

Light-induced conductivity enhancement that persists long after light exposure has ended, also known as persistent photoconductivity (PPC), has long been known to be a problem in ZnO-based ultraviolet photodetectors.[12] Historically, PPC in ZnO has largely been attributed to a metastable conductive oxygen vacancy state ($V_O \rightarrow V_O^{2+}$) induced by photoexcitation of trapped electrons at the surface of the material.[13] Recent investigations have suggested that, in addition to surface oxygen vacancies, PPC can also be partly attributed to charged zinc vacancies ($V_{Zn}^{2+}$) and interstitial defects, as well as stable and metastable defect complexes involving zinc vacancies and hydrogen impurities within the material bulk.[14] In this letter, we report on the electrical and structural characterization of thin-film ZnO MSM UV photodetectors subjected to a high fluence of 200 keV temperature-dependent proton irradiation.

To fabricate the ZnO MSM UV photodetector, we began with deposition of 1 µm of amorphous SiO$_2$ on a 525 µm (100) p-type 4-inch silicon wafer substrate using plasma-enhanced chemical vapor deposition (PECVD) system (PlasmaTherm Shuttlelock SLR 730) to electrically isolate the ZnO film from the effects of the Si substrate. Next, ~40 nm of ZnO was deposited via atomic layer deposition (ALD) at 150°C (Cambridge NanoTech Savannah S200). Thicknesses of both films were verified with ellipsometry. Next, contacts were formed using a standard lift-off procedure with 40 nm of evaporated Au. The exposed ZnO surface area is 0.151 mm$^2$ and the contact area is 0.386 mm$^2$, with interdigitated electrodes that are 500 µm long, 10 µm wide, and have an interelectrode spacing of 10 µm. Finally, a rectangular trench was etched around each device for electrical isolation. Figure 1 shows scanning-electron microscopy (SEM) images and cross-sectional schematics of the fabricated devices.

Devices were irradiated with 200 keV protons up to a fluence of $10^{16}$ protons/cm$^2$ on an ion implanter (Danfysik, Inc.) at the Ion Beam Materials Laboratory (IBML) in Los Alamos National Laboratory (LANL). Ion implantation profiles were generated using Stopping Range of Ion in Matter (SRIM). The substrate was angled 7° off from the proton beam to prevent channeling effects. During irradiation, devices were held either at low temperature (-25°C), room temperature (25°C), or high temperature (70°C) by heating or cooling the target stage (See Supplemental Information). To better simulate the effects of space-borne radiation on active devices, the detectors were biased with 1 V several times per minute during irradiation, as the presence of an electric field during irradiation is known to have a significant effect on the resulting damage profile.[1] Devices were annealed at room temperature for 6 months before electrical and microstructural characterization.

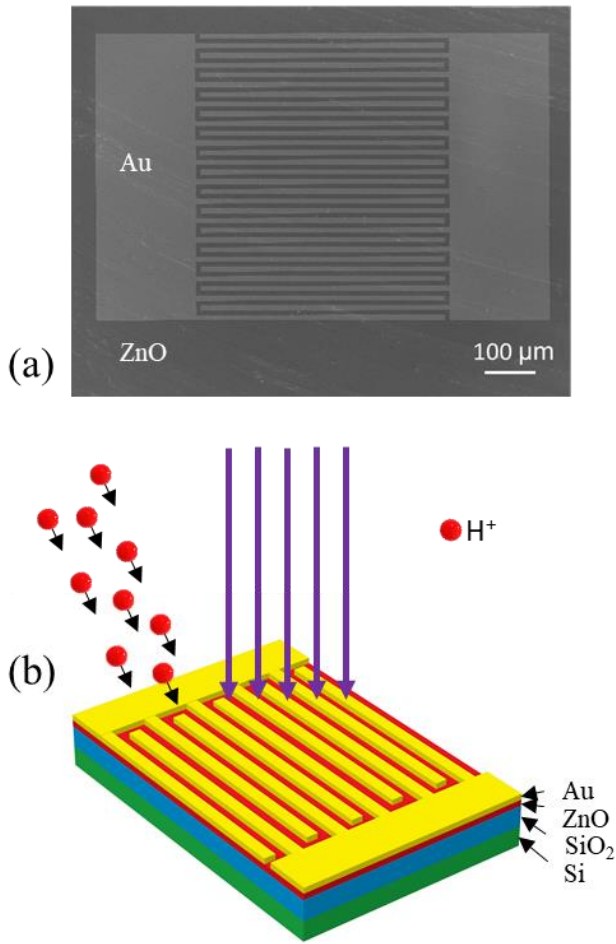

FIG. 1. (a) Top-view SEM of the ZnO MSM UV photodetector and cross-sectional schematics showing (b) proton irradiation and UV illumination of the device

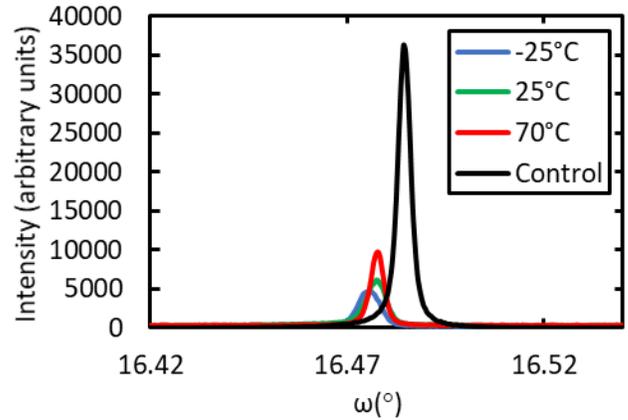

FIG. 2. Measured XRD rocking curve spectra and full-width half-maximums of the ZnO (100) peak for irradiation temperatures.

TABLE I. XRD Rocking Curve FWHMs for all irradiation temperatures

| Irradiation Temp. | FWHM (°) |
| --- | --- |
| -25°C | 0.0067 |
| 25°C | 0.0059 |
| 70°C | 0.0040 |
| Control | 0.0040 |

XRD (Philips X'Pert, copper Kα X-Ray source) rocking curves were used to investigate the crystal structure of the ZnO before irradiation and after irradiation at all temperatures. Figure 2 and Table 1 show the rocking curves and associated full-width half-maximums (FWHMs) for sample at various irradiation temperatures. XRD analysis of the films indicates a (100) preferred orientation, in line with previous reports of low-temperature ALD ZnO on glass.[15,16,17,18,19] This is in contrast to typical ZnO film growth, in which the (002) orientation is preferred. Surface migration is believed to be an important factor in c-axis-oriented growth, and so is inhibited during lower-temperature growths.[20] The control sample (100) peak is shifted ~0.5° to the right of its theoretical position indicating tensile strain in the as-deposited film. All irradiated devices show shifts of the ZnO (100) peak to the left relative to the control, which is indicative of strain relaxation with devices irradiated at lower temperatures exhibiting slightly larger shifts.

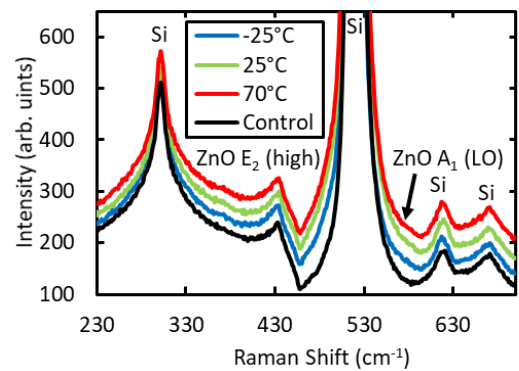

FIG. 3. Measured Raman spectra for irradiation temperatures.

Raman Spectroscopy (HORIBA Scientific LabRAM HR Evolution spectrometer, 532 nm laser) was also used to investigate the changes induced by proton irradiation. Figure 3 shows the Raman spectra for samples irradiated at all three temperatures, as well as the control sample. The peaks at 302 cm$^{-1}$, 528 cm$^{-1}$, 620 cm$^{-1}$, and 670 cm$^{-1}$ correspond to the silicon substrate,[21] and the peaks at 433 cm$^{-1}$ and ~573 cm$^{-1}$ are attributed to the ZnO $E_2$ (high) and $A_1$ longitudinal optical (LO) modes, respectively.[21,22] The $E_2$ (high) peak is shifted to the left for all samples relative to its theoretical position at 437 cm$^{-1}$, which is indicative of tensile strain in the as-deposited ZnO film.[21] For the irradiated samples, the $E_2$ peak is shifted towards its bulk position (~435 cm$^{-1}$ for the -25°C sample, ~434 cm$^{-1}$ for the 25°C and 70°C samples), which indicates a radiation-induced partial relaxation of as-deposited tensile strain, which is in good agreement with the results from XRD.[21,23] The 573 cm$^{-1}$ peak, which is associated with the presence of $(V_O)$ and $(Zn_i)$ is not distinct, likely for three reasons: because it is partially buried by the strong peak from the silicon substrate at 528 cm$^{-1}$, because the ZnO film is only 40 nm thick, and because the appearance of this peak in Raman spectra is suppressed by the presence of hydrogen, and, as indicated by the SRIM simulation, significant amounts of hydrogen were introduced by the proton irradiation.[21,24]

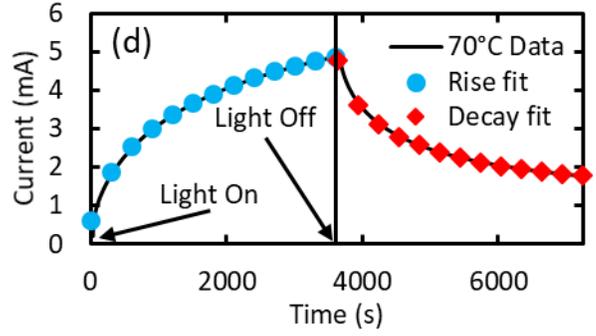

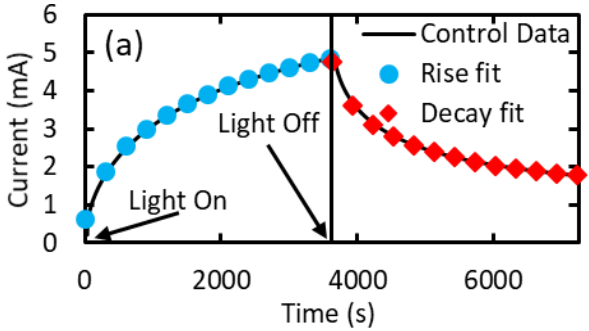

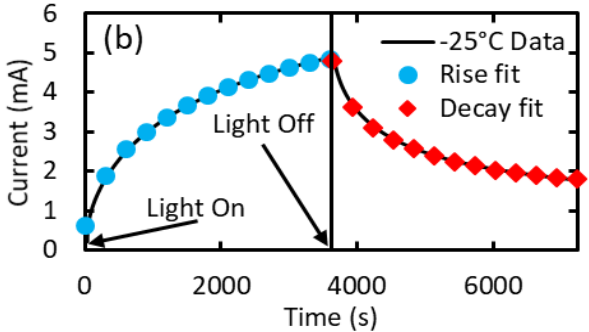

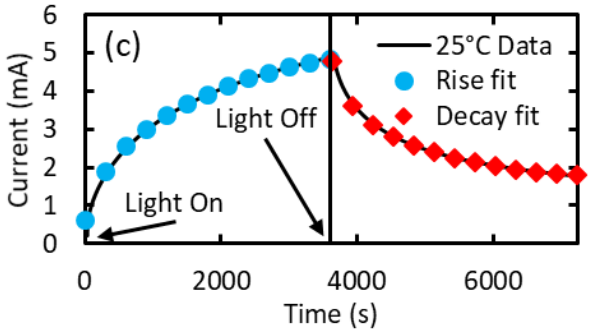

FIG. 4. Measured photocurrent transients and fits for (a) control, (b) low temperature, (c) room temperature, and (d) high temperature samples.

Table II. Photocurrent Rise Time Constants

| Irradiation Temp. | $\tau_1$ (s) | $\tau_2$ (s) |
|---|---|---|
| -25°C | 3391 | 270.9 |
| 25°C | 2855 | 245.5 |
| 70°C | 2596 | 230.9 |
| Control | 2034 | 234.2 |

Table III. Photocurrent Decay Time Constants

| Irradiation Temp. | $\tau_1$ (s) | $\tau_2$ (s) |
|---|---|---|
| -25°C | 4154 | 350 |
| 25°C | 3852 | 321.2 |
| 70°C | 2112 | 102.8 |
| Control | 1509 | 211.9 |

In addition to structural characterization, photocurrent vs. time measurements were taken to study the effects of radiation on device performance. Each device was biased at 1 V for 30 seconds, then illuminated with a 365 nm UV LED for one hour, after which the light was turned off and the photocurrent decay was observed.

It was found that both the photocurrent rise and decay were best modeled by a sum of exponentials, in agreement with results from literature.[25,26,27] Equation 1 was used to fit photocurrent rise data and equation 2 was used to fit photocurrent decay data, where $i$ is current, $t$ is time, $a$, $b$, and $c$ are fit constants, and $\tau_1$ and $\tau_2$ are time constants which correlate to activated defect relaxation phenomena. It was found that decreasing the temperature during irradiation substantially increased the value of both time constants during both photocurrent rise and fall, with some values for the -25°C samples being more than twice those of the control samples. Figure 4 displays the data and fits for all four sample conditions, and tables 2 and 3 display the time constants, respectively.

$$i = a*\left(1-e^{\left(-\frac{t}{\tau_1}\right)}\right) + b*\left(1-e^{\left(-\frac{t}{\tau_2}\right)}\right) + c \quad (1)$$

$$i = a*\left(e^{\left(-\frac{t}{\tau_1}\right)}\right) + b*\left(e^{\left(-\frac{t}{\tau_2}\right)}\right) + c \quad (2)$$

In summary, 50 nm ZnO MSM UV photodetectors were characterized electrically, structurally, and spectroscopically before and after being subjected to 200 keV proton irradiation up to a fluence of $10^{16}$ cm$^{-2}$ while held at different temperatures. XRD rocking curves and Raman spectra indicate significant increases in defect densities and partial relaxation of as-deposited tensile strain for all irradiated samples relative to the control, with the devices irradiated at lower temperatures experiencing the most damage. Photocurrent vs time measurements under 365 nm UV illumination showed significant increases in time constants as irradiation temperature was decreased, indicating that the temperature of ZnO devices during irradiation has a profound effect on dynamic annealing capability and therefore defect accumulation, significantly affecting long-term device performance. See Supplemental Information for the proton irradiation test setup and SRIM simulation.

This work was performed, in part, at the Center for Integrated Nanotechnologies, an Office of Science User Facility operated for the U.S. Department of Energy (DOE), Office of Science. Los Alamos National Laboratory, an affirmative action equal opportunity employer, is operated by Los Alamos National Security, LLC, for the National Nuclear Security administration of the U.S. Department of Energy under Contract No. DE-AC52-6NA25396. Fabrication and material characterization work were performed in part at the Stanford Nanofabrication Facility (SNF) and Stanford Nano Shared Facilities (SNSF). This material is based upon work supported by the U.S. Department of Energy, Office of Science, Office of Workforce Development for Teachers and Scientists, Office of Science Graduate Student Research (SCGSR) program. The SCGSR program is administered by the Oak Ridge Institute for Science and Education for the DOE under contract number DE-SC0014664.